\documentclass[twocolumn,showpacs,preprintnumbers,amsmath,amssymb]{revtex4}
\usepackage{graphicx}
\begin{document}
\title
{\large{\bf  Spontaneously generated X-shaped light bullets}}
\author{P. Di Trapani,$^1$ G. Valiulis,$^2$
A. Piskarskas,$^2$, O. Jedrkiewicz,$^1$ J. Trull$^1$ , C.
Conti,$^3$ S. Trillo,$^{3,4}$}
%\altaffiliation[Also at ]{Dept. Fisica i Enginyeria Nuclear, UPC Terrassa, Spain}

\affiliation{$^1$INFM and Department of Chemical, Physical and  Mathematical
Sciences, University of Insubria, Via Valleggio 11, 22100 Como, Italy}

\affiliation{$^2$ Department of Quantum Electronics, Vilnius
University, Sauletekio al. 9, bldg. 3, LT-2040 Vilnius,
Lithuania}

\affiliation{$^3$ Istituto Nazionale di Fisica della Materia (INFM)-RM3,
Via della Vasca Navale 84, 00146 Roma, Italy}

\affiliation{$^4$ Department of Engineering,
University of Ferrara, Via Saragat 1, 44100 Ferrara, Italy}

\date{\today}

\begin{abstract}
We observe the formation of an intense optical wavepacket fully
localized in all dimensions,
i.e. both longitudinally (in time) and in the transverse plane,
with an extension of a few tens of fsec and microns, respectively.
Our measurements show that the self-trapped wave is a X-shaped
light
bullet spontaneously generated from a standard laser wavepacket via
the nonlinear material response (i.e.,
second-harmonic generation), which extend the soliton concept to a new realm,
where the main hump coexists with conical tails which reflect
the symmetry of linear dispersion relationship.
\end{abstract}

\maketitle
\begin{figure}
\includegraphics[width=6cm]{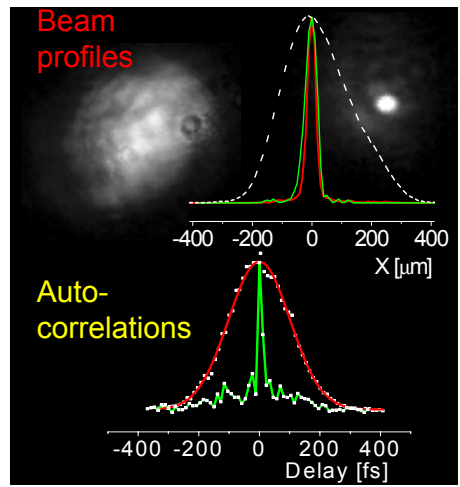}
\caption{Top: output transverse beam profiles
in the linear (white) and nonlinear (green: SH; red: FF) regime.
The diffracted (left) and localized (right) spots recorded
at the output of the crystal are also shown.
Bottom: temporal autocorrelation traces (collinear technique)
showing the transition from the linear regime (red) to the
compressed WP (green).}
\label{exp1} \end{figure}
\begin{figure}
\includegraphics[width=6cm]{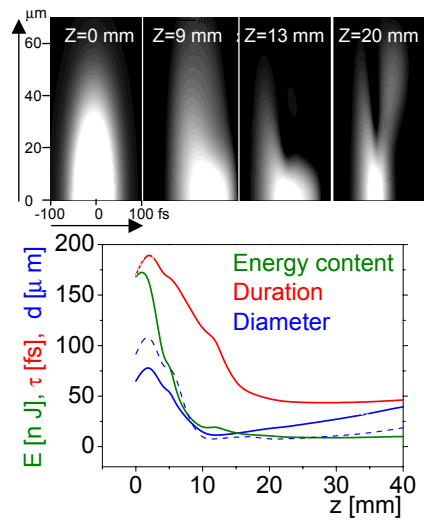}
\caption{Numerical simulation of the localization
process observed in our LBO sample
[from Eqs.~(\ref{SHGeqs}) with
$k''(\omega_0)=0.016~{\rm ps}^2$/m, $k''(2\omega_0)=0.089~{\rm ps}^2$/m,
$\delta V= 45$ ps/m, $\chi = 7 \times 10^{-5}$ W$^{-1/2}$
($d_{eff}=0.85$ pm/V)]
from an input gaussian WP with $45~\mu$m waist, $170$ fs duration,
and $0.26~\mu$J energy.
Top: snapshots showing the FF beam in the
$t-x$ plane. Bottom: evolution of the beam
diameter, duration, and energy content of main localized hump.
%1) Energy=, I=30GW/cm^2, dur=170fs, diam=65mkr, diam_at_waist=40mkr,
%>focused 7.1mm before cryst.
%>2) Energy=0.74mkJ, I=30GW/cm^2, dur=240fs, diam=92mkr, diam_at_waist=56.5mkr,
%>focused 14mm before cryst.
The dashed line refers to more energetic (e.g., slightly wider and longer)
input WPs for which even the small residual diffraction (solid line)
is removed.
In this case, however, the reshaping leads to higher peak intensity
close to damage threshold of LBO.}
\label{numerics}
\end{figure}
\begin{figure}
%\hspace{-1.5cm}
\includegraphics[width=8cm]{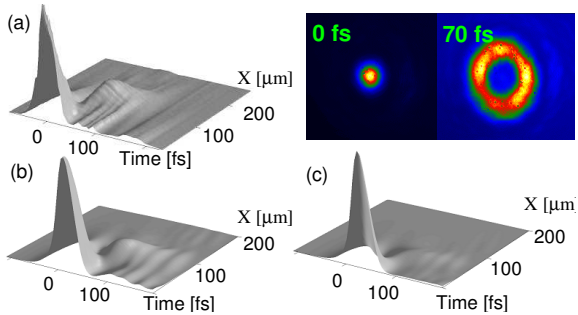}
\caption{(a) Output spatio-temporal intensity profile, as measured
in air, $5$ mm from the crystal output face. Insets: transverse
intensity pattern in  $(x,y)$ plane measured at peak ($t=0$ fs)
and with $t=70$ fs delay. (b) As in (a), numerical result from
Eqs. (2). (c) As in (b) calculated right on crystal output.}
\label{exp2} \end{figure}

Defeating the natural spreading of a wavepacket (WP) is a universal
and challenging task in any physical context involving wave
propagation.
Ideal particle-like behavior of WPs is demanded
in applications, such as microscopy, tomography, laser-induced
particle acceleration, ultrasound medical diagnostics, Bose-Einstein
condensation,  volume optical-data storage, optical interconnects,
and those encompassing long-distance or high-resolution signal
transmission.
The quest for light WPs that are both invariant (upon propagation)
and sufficently localized in all dimensions (3D, i.e.,
both transversally and longitudinally or in time)
against spreading "forces" exerted by diffraction and material group-velocity
dispersion (GVD, $k''=d^2 k/d \omega^2|_{\omega_0}$) has motivated
long-standing studies, which have followed different
strategies in the {\em linear}
\cite{durnin87,theoryXwave,lg92,saari97,mugnai00}
and {\em nonlinear} \cite{bullet,wise} regime, respectively.
%%% FIGURE 1 %%%
\begin{figure}
\includegraphics[width=8cm]{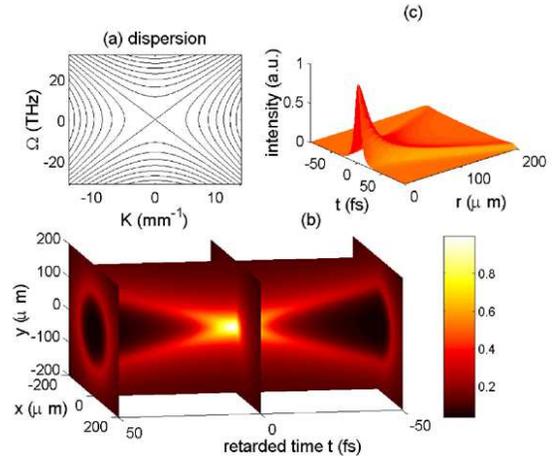}
\caption{Features of normally dispersive media ($k''>0$): (a)
spatio-temporal dispersion relation associated with paraxial wave
equation~(\ref{parax}); Different representations [(b) sections of
color isointensity surfaces $|E(x,y,t)|^2=$const.; (c) intensity
vs. $r,t$] of a non-diffractive, non-dispersive linear X-wave
($\Delta=10$ fs, $k''=0.02$ ps$^2$/m, $k_0=10^7$ m$^{-1}$).}
\label{linear} \end{figure}
%%%%%%%%%%%%
\indent In the {\em linear} case, to counteract material
(intrinsic) GVD, one can exploit the angular dispersion (i.e.,
dependence of propagation angle on frequency) that stems from a
proper WP shape. The prototype of such WPs is the X-wave
\cite{theoryXwave}, a non-monochromatic, yet non-dispersive,
superposition of non-diffracting cylindrically symmetric Bessel
$J_0$ (so-called conical or Durnin \cite{durnin87}) beams,
experimentally tested in acoustics \cite{lg92}, optics
\cite{saari97} and microwave antennae \cite{mugnai00}.
Importantly, in the relevant case of WPs with relatively narrow
spectral content both temporally (around carrier frequency
$\omega_0$) and spatially (around propagation direction $z$, i.e.
paraxial WPs), X-waves require {\em normally} dispersive media
($k''>0$). In this case, a WP with disturbance $E(r,t,z) \exp(i
k_0 z -i \omega_0 T)$ ($k_0 \equiv k(\omega_0)$, $r^2 \equiv
x^2+y^2$), has a slowly-varying envelope $E=E(r,t,z)$ obeying the
standard wave equation
\begin{equation} \label{parax}
{\displaystyle \hat{L}(\omega_0)~E=0\;;\;
\hat{L}(\omega_0) \equiv i \partial_{z} + \frac{1}{2k_0}
\nabla^2_{\perp} - \frac{k''}{2} \partial_{tt}^2}.
\end{equation}
%where $\nabla^2_{\perp}=\partial_x^2+\partial_y^2$ is the transverse
Laplacian,
where $\nabla^2_{\perp}=\partial_{rr}^2+r^{-1} \partial_r$ is the transverse
Laplacian, and we limit our attention to luminal WPs traveling at
light group-velocity
$1/k'=dk/d \omega|_{\omega_0}^{-1}$ by introducing
the retarded time $t=T-k' z$ in the WP barycentre frame.
Propagation-invariant waves $E(r,t,z)=E(r,t,z=0) \exp(i \beta z)$
can be achieved whenever their input spatio-temporal spectra $E(K,\Omega,z=0)$
lie along the characteristics of the dispersion relationship
$k'' \Omega^2/2-K^2/(2k_0)=\beta$, which follows from
Eq.~(\ref{parax}) in Fourier space
$(K,\Omega)$ ($K$ is the transverse wavevector related to cone angle
with $z$-axis
$\theta \simeq \sin \theta = K/k_0$, and $\Omega=\omega-\omega_0$).
In the normal GVD regime ($k''>0$) these curves, displayed in
Fig.~\ref{linear}(a),
reflect the hyperbolic nature of the wave equation~(\ref{parax})
and show the common asymptotic spectral X-shape associated with the
lines $K=\pm 2k_0 k''
\Omega$ ($\beta=0$). When the spectral components lying along such X
are superimposed
coherently (in phase), the field in the physical space $(r,t)$
also retains a propagation invariant X-shape, as entailed e.g. by the
exact solution
$E={\rm Re} \{[(\Delta-it)^2 + k_0 k'' r^2]^{-1/2} \}$
of Eq.~(\ref{parax}) shown in Fig.~\ref{linear}(b-c).
Here $\Delta$ represents duration of the X-wave central hump.
Main features of these waves are the conical (clepsydra) 3D structure
and the slow spatial decay ($1/r$ characteristic of $J_0$ components
\cite{durnin87}), displayed by
Fig.~\ref{linear}(b) and (c), respectively.
\newline\indent
Conversely, in the {\em nonlinear} (high intensity) regime,
non-spreading WPs exploit the idea that, in self-focusing media,
the nonlinear wavefront curvature can simultaneously balance the
curvature due to diffraction and to GVD, combining features of
spatial \cite{soliton} and temporal \cite{temporal} solitons to
form a bell-shaped 3D-localized WP $E(r,t)$, so-called
light-bullet \cite{bullet}. In sharp contrast with X-waves, such
compensation strictly requires {\em anomalous} GVD ($k''<0$)
\cite{bullet}, thus implying that along the WP tails, where
Eq.~(1) still holds true, space $r$ and time $t$ play the same
role giving rise to strong WP localization \cite{math}. Stable
trapping, however, has been observed only in setting of reduced
dimensionality  (2D) \cite{wise,eis01}, including also other
contexts, e.g. spin \cite{spin} or atomic waves \cite{BEC1} (3D
kinetic energy and atom-atom attractive interactions act exactly
as diffraction-GVD and self-focusing, respectively) where similar
trapping mechanisms hold true.
\newline\indent
In this work, we outclass the two approaches by demonstrating that
{\em space-time localization in the normal GVD regime becomes accessible in
the nonlinear regime}. Trapping is accomplished by mutual balance
of intrinsic, shape-induced, and nonlinear contributions in
a new type of WP, namely a nonlinear X-wave,
which permits to get over two limitations at once.
First, in contrast with linear X-waves,
whose observation requires non-trivial input beam shaping \cite{lg92,saari97},
our experiment reveals a remarkable "mode-locking" process
that, starting from a conventional (gaussian) laser WP,
spontaneously performs the reshaping into a localized X-shaped WP.
Second, we believe this to be the first genuine nonlinear trapping
in full-dimensional 3D physical space, since to date
material and/or instability limitations \cite{bullet,wise} have rendered
the observation of light 3D bullets elusive.
\newline\indent
Figure~\ref{exp1} describes the strong localization features
observed after propagation in a $22$ mm long sample of lithium
triborate (LBO) $\chi^{(2)}$ crystal. At the input we launch a
laser WP at fundamental frequency (FF) $\omega_0=2 \pi
c/\lambda_0$, $\lambda_0=1060$ nm, with gaussian profile in both
$t$ (with FWHM duration in the $100-200$ fs range ) and $r$
($45~\mu$m FWHM at waist, located few mm before the crystal so
that the input beam is slightly diverging). The LBO crystal is
tuned for generation of optical second-harmonic (SH) in the regime
of relatively large positive phase mismatch $\Delta
k=2k(\omega_0)-k(2\omega_0)=30~cm^{-1}$ or effective self-focusing
for the FF beam \cite{reviewSHG}. When the input energy exceeds
about $0.25$ $\mu$J, mutually trapped localized WPs at FF and SH
are observed (see Fig.~\ref{exp1}). Time-integrated measurements
of spatial profiles (Fig.~\ref{exp1}, top) indicate that
diffraction is fully defeated to yield a spatial soliton-like beam
\cite{soliton}, while temporal autocorrelations reveal a single
pulse which is strongly compressed (down to $\sim 20$ fs, see
Fig.~\ref{exp1}, bottom). Remarkably nonlinear mixing balances the
two highest-order dispersive effects, namely the tendency of FF
and SH to walk-off due to group-velocity mismatch (GVM) $\delta
V=k'(2 \omega_0)-k'(\omega_0)$ and GVD, which act in the linear
regime over characteristic length scales $50$ and $2.5$ times
shorter than the crystal length, respectively. Relying on
nonlinear scenarios known to date, the present result is
unexpected. In fact, the normal GVD and the strong GVM of LBO do
not allow an explanation in terms of light bullets, whereas the
dynamics of self-focusing with normal GVD is dominated by pulse
splitting without envisaging localization whatsoever
\cite{splitting}.
%%% FIGURE 2 %%%
%%%%%%%%%%%%
\newline\indent In order to get a deeper understanding, we performed a
set of numerical experiments by integrating the well-known model
\cite{reviewSHG}, which generalizes Eq.~(\ref{parax}) to nonlinear
coupling of envelopes $E_m(r,t,z)$ ($m=1,2$, carrier $m \omega_0$)
\begin{equation}
\label{SHGeqs}
\begin{array}{l}
{\displaystyle \hat{L}(\omega_0)~E_1 + \chi E_2 E_1^* \exp(-i \Delta
k z) =  0,} \\
{\displaystyle \hat{L}(2 \omega_0)~E_2 + \delta V \partial_t E_2 +
  \chi E_1^2 \exp(i\Delta k z) = 0,}
\end{array}
\end{equation}
%\newline\indent
In Fig.~\ref{numerics} (top) we show snapshots of the space-time
evolution. It is clear that, in the first stage ($z=10-15$ mm),
the generation of light at SH is accompanied by strong
self-focusing, in turn inducing pulse compression (in spite of the
normal GVD that would cause temporal broadening of plane-waves).
The WP becomes asymmetric since self-focusing is stronger where
the SH, which lags behind because of GVM, tends to be. More
importantly, during this process the WP undergoes a strong
reshaping, where a large fraction of energy is radiated off-axis
to form the tails of a conical wave (see snapshot at $z=20$ mm).
After this stage, however, the collapse stops and the WP
propagates (locked with the SH) with immutable shape and nearly
constant energy, duration, and size.
\newline\indent
By changing the parameters in Eqs.~(\ref{SHGeqs}) we can conclude
that: (i) nonlinearity is the key element that drives the
reshaping and holds the WP together. In fact, by switching it off
after the transient (i.e., $\chi=0$ for $z > 20$ mm), the WP
exhibits strong diffraction and extremely fast FF-SH walk-off. The
reshaped WP can by no means be considered a linear X-wave; (ii) in
our crystal, GVM is the dominant dispersion term affecting the
asymptotic duration and width of the localized WP (both decrease
for smaller GVM). However, (symmetric) X-shaped WPs are formed
also in the ideal case of vanishing GVM, provided that GVD is
normal; (iii) the WP reshaping strongly affects the phase
modulation process, which reveals the contribution of an effective
anomalous GVD; (iv) by including additional cubic nonlinearities
\cite{reviewSHG}, the phenomenon remains qualitatively unchanged.
\newline\indent
A more rigorous ground for explaining the dramatic reshaping shown
in Fig.~\ref{numerics} is offered by the investigation of two
different aspects, both implicitly accounted for by
Eqs.~(\ref{SHGeqs}): (i) the stability of continuous plane-waves;
(ii) the existence of nonlinear X-shaped eigensolutions. Though
somewhat idealized, the results of this analysis allow gathering
the two basic features of the spatio-temporal dynamics, namely the
transient reshaping of the input gaussian-like WP followed by the
quasi-stationary regime.
\newline\indent
The linear stability analysis of $z$-invariant solutions of
Eqs.~(\ref{SHGeqs}) with ideally vanishing spatio-temporal
spectral width (i.e., continuous plane waves) always reveal the
presence of exponentially growing weak (up to noise level)
perturbations with definite frequency $K,\Omega$. However, the
instability features reflect the symmetry of the linear wave
equation, thus being qualitatively different in the  normal and
anomalous GVD regime, respectively. While anomalous GVD leads to
narrow bandwidth features as in conventional spatial or temporal
modulational instability \cite{reviewSHG}, the hyperbolic
structure of the diffraction-dispersion operator in the normal GVD
regime of our experiment leads to exponential amplification of
conical wave perturbations (Bessel $J_0$ beams) with frequencies
basically approaching the asymptotes of Fig.~\ref{linear}(a)
\cite{conicalSHG}. When growing spontaneously from noise, the
amplified components in the virtual infinite bandwidth (in reality
limited by non-paraxiality and higher-order dispersion not
accounted for in Eqs.~(1-2)) leads to colored conical emission.
Our calculations show that the phenomenon persists under dynamical
conditions (unseeded SH generation) and when pumped by short-pulse
narrow-beam inputs, as in our experiment. In the latter case, the
amplification of proper frequency components of the WP preserves
the mutual phase coherence, thus acting as a trigger which drives
the transformation of the WP into the X-wave shown in
Fig.~\ref{numerics}. In other words, it is the conical instability
that probes the symmetry of the underlying linear system in
amplifying those components which allow both diffraction and
dispersion of the whole WP to be removed.
\newline\indent
A second, strong argument in favour of the nearly asymptotic
character of the evolution shown in Figs.~\ref{numerics}, is the
existence of stationary localized solutions of
Eqs.~(\ref{SHGeqs}). We have recently shown, indeed, that the
natural propagation-invariant [i.e., with dependence $E_m(r,t)
\exp(i m \beta z)$, $m=1,2$] localized eigensolutions of SH
generation process in the normal GVD regime (either with or
without GVM) are indeed nonlinear X waves \cite{NLXW}. These waves
are similar to those shown in Fig.~\ref{linear}(b-c) except for
the fact that their peak intensity is related to their duration
and width through the nonlinearity.
%%% FIGURE 3 %%%
%%%%%%%%%%%%
\indent Since the measurements in Fig.~\ref{exp1} give information
only about the WP central hump, in order to confirm experimentally
that the observed strong focusing and compression dynamics is
indeed driven by the formation of an X-wave, we have performed
different additional measurements. First, we have characterised
the propagation of the WP in air after the LBO nonlinear crystal.
We observe that the beam diffracts less than a gaussian beam of
the same width (the divergence angle is $2.5$ times less), and
this net sub-gaussian diffraction witnesses a Bessel-like feature
of the WP. Moreover, temporal broadening to $150$ fs after only
$10$ cm of propagation in air is an indication that the WP
develops strong angular dispersion. Both observations are in good
quantitative agreement with data obtained numerically from
Eqs.~(\ref{SHGeqs}). However, it is only the tomography of the
output WP, i.e. mapping the WP intensity in space and time, that
can give direct unequivocal evidence for the formation of an
X-wave. To this end we have developed a new technique based on an
ultrafast nonlinear gating, or a scanning cross-correlation
technique, realized by frequency mixing the WP under investigation
with a $20$ fs, high contrast, steep front, probe pulse in the
visible, which is uniform over a large (few mm$^2$) area (the
details of the set up will be presented elsewhere). Thanks to the
high quality of the probe and the use of a very thin ($20~\mu$m)
BBO mixing crystal, the apparatus has high temporal resolution.
The intensity map of the output WP is reported in
Fig.~\ref{exp2}(a). The insets also show the measured beam profile
in the transverse plane at time $t=0$ (peak) and $t=70$ fs (far
from peak), respectively. The measured profile clearly shows the
features of an X-wave with a conical structure, which emanates
from a strongly localized central spatio-temporal hump. Unlike
conventional pulse splitting \cite{splitting}, here splitting
occurs only sufficiently off-axis ($x \sim 100 \mu$m) in the WP
low-intensity portion. For comparison we also report in
Fig.~\ref{exp2}(b) the profile calculated from Eqs.~(2) under the
same conditions, the agreement being excellent. The fringe-like
structure that appears for large delays in both Fig.~\ref{exp2}(a)
and (b) is due to $5$ mm of free-space propagation in air outside
the crystal. Although the calculated profile on the output face of
the crystal indeed shows that such fringes disappear [see
Fig.~\ref{exp2}(c)], measurement with perfect imaging on the
output LBO face reveals saturation of the mixing process due to a
too intense peak.
%%% FIGURE 4 %%%
%%%%%%%%%%%%
\newline\indent In summary, we have reported the first evidence
that the natural 3D (temporal and spatial) spreading of a focused
ultrashort wavepacket can be balanced in transparent materials at
high intensity. The underlying mechanism is the {\em spontaneous}
formation of a X-wave characterized by an intense (i.e.,
nonlinear) central hump self-trapped through mutual balance with
(essentially linear) dispersive contributions associated with
coexisting slowly decaying conical tails. While our experiment is
carried out by exploiting self-focusing nonlinearities arising
from quadratic nonlinearity, we envisage the general role that
self-trapping mediated by nonlinear X-waves can have for a wide
class of materials and applications encompassing centrosymmetric
optical (Kerr) media \cite{bullet}, Bose-Einstein condensation
\cite{BEC1}, and acoustics \cite{lg92}.
\newline\indent
We acknowledge support from MIUR (PRIN and FIRB projects), Unesco
UVO-ROSTE (contract 875.586.2), Lithuanian Science and Studies
Foundation (grant T-491), Secretaria de Estado y Universidades in
Spain, and Fondazione Tronchetti Provera in Italy. We are grateful
to the technical assistance of Light Conversion Ltd.

\end{document}